# Direct simulation of aerodynamic entrainment with inter-particle cohesions

Shuming Jia [1], Zhengshi Wang [1, a)]

[1] College of Natural Resources and Environment, Northwest A&F University, Yangling, Shaanxi, 712100, China.

[a)] Corresponding author: Zhengshi Wang (wangzs2020@nwafu.edu.cn)

Running Head: Direct simulation of aerodynamic entrainment

**Abstract**

Aerodynamic entrainment acts as the pioneer of saltation movement and is critical for understanding the development of aeolian phenomena. Here we performed direct numerical simulations on the aerodynamic lifting of surface particles on a random arranged sediment bed using the discrete element method, and provided evidence that particles do not leave the bed vertically with speeds equal to that needed to reach a height of one grain diameter, as people widely accepted, but with much smaller angles toward the bed surface and much larger velocities. The entrainment rate does not exhibit a quadratic dependence with the shear velocity, but a nearly linear one, because the velocity of entrained particle increases linearly with shear velocity. Moreover, inter-particle cohesion increases the fluid entrainment threshold significantly, but has no effect on the entrainment rate, as well as the take-off velocity and angle distributions of entrained particles, because the collisions between entrained and bed particles destroy the inter-particle bonds. The completely entrainment scheme is given, which provides a credible approach to explore the development and evolution of aeolian transports.

**Keywords**: Aerodynamic entrainment; discrete element method (DEM); inter-particle cohesion; probability density function (PDF); particle trajectory.

## 1. Introduction

The wind-blown sand (or snow) movement and dust transportation are typically initiated by the bombardment of several aerodynamic entrainment particles (Kok *et al.*, 2012; Pähtz *et al.*, 2020; Shao, 2008), in which the aerodynamic entrainment describes the lifting of surface particles directly by fluid drag. Generally, the aerodynamic entrainment within a steady state saltation is believed to be negligible (Anderson and Haff, 1991; Kok *et al.*, 2012). Recent models derived with the help of DEM simulations also shown that neglecting aerodynamic entrainment can reproduce the experimental mass flux exactly (Pähtz and Durán, 2020). However, this assumption has been challenged in drifting snow both by measurements (Doorschot and Lehning, 2002) and model predictions (Nemoto and Nishimura, 2004). Recent experiments also argue that aerodynamic entrainment during unsaturated sand transport is significant (Li *et al.*, 2020).

Plenty of investigations have been conducted to explore how the first particle enters into the air by fluid drag, including experimental observations (Agudo and Wierschem, 2012; Charru *et al.*, 2007; Dey and Papanicolaou, 2008; Dwivedi *et al.*, 2011; Hofland *et al.*, 2005; Houssais *et al.*, 2015) and numerical simulations (Agudo *et al.*, 2017; Buffington and Montgomery, 1997; Lee and Balachandar, 2012; Pan and Banerjee, 1997; Soldati and Marchioli, 2009; Vinkovic *et al.*, 2011; Vowinckel *et al.*, 2016). These works generally concentrate on the onset of single or several particles on a regular bed, while the entire aerodynamic entrainment process over a random arranged sediment bed with multi-body contact, friction, sliding, separation and inter-particle cohesion has been rarely involved. Nevertheless, a linear scaling for the entrainment rate as a function of the fluid shear stress, obtained from the momentum balance at the surface, originally presented by Anderson and Haff (1988), is widely used in both wind-blown sand and snow models (Anderson and Haff, 1988; Huang and Wang, 2016; Nemoto and Nishimura, 2004; Zwaaftink *et al.*, 2014). At the same time, most present take-off velocity and angle distributions used in the entrainment model are based on mathematical statistics or artificial assumptions (i.e., with uniform vertical velocity equals to that needed to reach a height of one grain diameter (Anderson and Haff, 1988; Huang and Wang, 2016; Nemoto and Nishimura, 2004; Zwaaftink *et al.*, 2014)), or just initiated the saltation movement by releasing numbers of random pioneer particles (Dupont *et al.*, 2013; Huang and Shi, 2017), due to the lack of valid measurements or predictions (Shao, 2008). In this way, it is almost impossible to reproduce the exact development process (i.e., development time, saturation distance and unsaturated mass flux) of aeolian movements.

In this work, a realistic particle bed is established using the DEM method, and the entire process of a fluid-borne particle from the resting state to that can be counted as an entrained particle is performed. The energy accumulations of moving particles and the collisions with other particles during entrainment are analyzed detailed to reveal the entrainment mechanism of surface particles, and the widely used assumptions in aerodynamic entrainment scheme is examined. Finally, a completely entrainment scheme is summarized based on the DEM simulations, which provides a practical way to explore the development process of aeolian movements.

## 2. Model and Method

### 2.1 Discrete element method of particle motion

The behaviors of particle are simulated using the discrete element method (DEM), which updates the particle information (velocity and position) according to the applied contact force from adjacent particles and body force due to potential field (e.g., gravitational field):

$$m\frac{d\mathbf{v}_p}{dt} = \mathbf{F}_C + \mathbf{F}_D + \mathbf{G} \tag{1}$$

$$I\frac{d\boldsymbol{\omega}_p}{dt} = \mathbf{M}_C + \mathbf{M}_E \tag{2}$$

where $m = \pi d_p^3 \rho_p / 6$ is the particle mass, in which $d_p$ and $\rho_p$ are, respectively, the diameter and mass density of the particle, $\mathbf{v}_p$ and $\boldsymbol{\omega}_p$ are, respectively, the velocity and angular velocity of the particle, $t$ is time, $\mathbf{F}_D$ is the fluid drag force, $\mathbf{G} = m\mathbf{g}$ is the gravity with $\mathbf{g}$ the gravitational acceleration vector, and $\mathbf{F}_C = \sum_{i=1}^{n} \mathbf{F}_i$ is the resultant contact force by traversing all $n$ contacts, in which $\mathbf{F}_i$ is the contact force of the $i$th contact, which can be written as:

$$\mathbf{F}_i = \mathbf{F}_n + \mathbf{F}_s \tag{3}$$

where $\mathbf{F}_n$ and $\mathbf{F}_s$ are, respectively, the normal and frictional forces of the contact partner, which can be expressed as:

$$\mathbf{F}_n = \begin{cases} -\left(k_n \delta_n + m\gamma_n \dot{\delta}_n\right)\mathbf{n} & \begin{cases} \mathbf{bonded}: \left(-k_n \delta_n \leq F_\Phi\right) \\ \mathbf{unbonded}: \quad \left(\delta_n \geq 0\right) \end{cases} \\ 0 & others \end{cases} \quad (4)$$

$$\mathbf{F}_s = \begin{cases} \begin{cases} \mathbf{bonded}: \mathbf{F}_s^0 - k_s \boldsymbol{\delta}_s - m\gamma_s \dot{\boldsymbol{\delta}}_s & \left|\mathbf{F}_s^0 - k_s \boldsymbol{\delta}_s\right| \leq F_\Phi \\ \mathbf{unbonded}: \min\left(\mathbf{F}_s^0 - k_s \boldsymbol{\delta}_s - m\gamma_s \dot{\boldsymbol{\delta}}_s, -\mu|\mathbf{F}_n|\frac{\dot{\boldsymbol{\delta}}_s}{|\dot{\boldsymbol{\delta}}_s|}\right) & \delta_n \geq 0 \end{cases} \\ 0 & others \end{cases} \quad (5)$$

where $k_n = \pi d_p E/4$ and $k_s = k_n/k^*$ are, respectively, the normal and tangential contact stiffness, in which $E$ is the elasticity module, and $k^*$ is the normal-to-shear stiffness ratio, $\delta_n$ and $\Delta\delta_s = |\dot{\boldsymbol{\delta}}_s|\Delta t$ are, respectively, the normal overlap and tangential displacement increment, with the dot mark represents the time derivative, and $\Delta t$ is the time step controlled by $E$, $\mathbf{n}$ is the normal unit vector that point to the mass center of contact particle, $\gamma_n = \beta_n\sqrt{mk_n}$ and $\gamma_s = \beta_s\sqrt{mk_s}$ are, respectively, the normal and tangential viscous damping coefficient (Cundall, 1987), and $\mathbf{F}_s^0$ is the accumulated tangential force at the beginning of current time step. The inter-particle cohesion is included by applying the bond model (Comola *et al.*, 2019; Comola *et al.*, 2017; Gaume *et al.*, 2016), and the bond will become invalid either the normal or tangential force exceeds the given limitation $F_\Phi$ (Comola *et al.*, 2019), which is related to the material properties (i.e., the tensile and shear strength).

Besides, $\mathbf{M}_C = \sum \mathbf{F}_s \times \mathbf{r}$ and $\mathbf{M}_E$ are, respectively, the contact and applied external moment acting on the particle, in which $\mathbf{r}$ is the corresponding arm of force. That is to say, the moment $\mathbf{M}_C$ equals to the product of the applied force whose point of action deviated from the particle mass center and corresponding torque, and $I = md_p^2/10$ is the moment of inertia.

## 2.2 Fluid drag force model

The focus of this topic is the complex multi-body contact, separation and friction among particles; thus, the exact flow field integration may be unpractical. Till now, the calculation of $\mathbf{F}_D$ acting on a part exposed particle without resolving the flow field is still one of the most challenge issues. Following (Shih and Diplas, 2019), we firstly define the exposure $e$ as:

$$e = A_e / A_t \tag{6}$$

where $A_e$ and $A_t$ are, respectively, the exposed and total particle area projected upon a plane perpendicular to the flow.

Typically, a non-exposed particle ($e=0$) corresponds to a zero fluid drag force and the fluid drag force $\hat{\mathbf{F}}_D$ for a fully exposed particle ($e=1$) can be reasonable approximated by (Yamamoto et al., 2001):

$$\hat{\mathbf{F}}_D = m \frac{\mathbf{V}_r}{T_p} f(Re_p) \tag{7}$$

where $T_p = \rho_p d_p^2 / (18 \rho_f \nu) \sim 1.0e-2$ is the relaxation time, in which $\rho_f$ and $\nu$ are, respectively, the density and kinematic viscosity of the air, $\mathbf{V}_r = \mathbf{v}_f - \mathbf{v}_p$ is the relative velocity between fluid and particle, and $f(Re_p)$ is a function of particle Reynolds number $Re_p = d_p |\mathbf{V}_r| / \nu$ and can be written as (Clift et al., 1978):

$$\begin{cases} f(Re_p) = 1 & (Re_p < 1) \\ f(Re_p) = 1 + 0.15 \operatorname{Re}_p^{0.687} & (Re_p \geq 1) \end{cases} \tag{8}$$

For intermediate states ($0 < e < 1$), Shih and Diplas (2019) have shown that the fluid drag force is basically linear increased with the exposure. In this way, the fluid drag force can be written as:

$$\mathbf{F}_D = e\hat{\mathbf{F}}_D \qquad (0 \leq e \leq 1) \tag{9}$$

For a part exposed particle ($0 < e < 1$), the point of action of the fluid drag force is set as the middle height of the exposed portion. Here, the lift force is neglected due to its smaller order of magnitude than gravity and drag force (Shih and Diplas, 2019).

*2.3 Wind profile*

For aerodynamic entrainment of sediment particles, evidence has shown that very-large-scale flow structures are responsible for the initial rolling motion of bed particles (Cameron *et al.*, 2020). However, with the entrainment criterion defined in Sec. 2.4, the 'response time' of entrained particles should be much larger than that of the initial rolling motion. Thus, the mean flow velocity is adopted in this work.

The wind field contains three components (i.e., $\mathbf{v}_f = (u_x, 0, 0)$), in which the mean stream-wise component $u_x$ can be derived from the standard logarithmic law (Kok *et al.*, 2012):

$$u_x(z) = \frac{u_*}{\kappa} \ln\left(\frac{z}{z_0}\right) \tag{10}$$

where $\kappa$ is the Karman constant, $u_*$ is the fluid shear velocity, $z$ is the altitude above the bed surface ($H_{bed}$), and $z_0 = d_p/30$ is the aerodynamic roughness (Bagnold, 1941). Although a high viscous (laminar) layer may appear at the near surface, the rugged particle surface can enhance the turbulent mixing and vanish the viscous layer (Kok *et al.*, 2012). The parameter values are listed in Table 1.

*2.4 Model settings and simulation procedure*

The computational domain are $L_x \times L_y \times L_z = 40d_p \times 20d_p \times 60d_p$, with periodic boundary conditions along stream-wise ($x$) and span-wise ($y$) directions, and the reflective boundary condition is adopted at top and bottom. The simulation procedures are as following:

(1) Totally 8000 particles are created in the domain with random positions. All particles conduct a free-falling movement to construct a sediment bed. The criterion for the calm bed is $v_{p\max}/\sqrt{gd_p} < 2.0e-3$, where $v_{p\max}$ is the maximum particle velocity.

(2) Remove the particles whose top is greater than $H_0(=12d_p)$ to form a flat surface, and re-calm the bed. Repeat this step twice to guarantee a steady state of the sediment bed. The porosity $p$ of the established bed is approximately 0.5.

(3) Determine the bed height $H_{bed}$ at which the mean wind velocity reduces to zero, as shown in Figure 1. Considering the porosity of sediment bed, $H_{bed}$ can be read as $H_{bed} = H_0 - pd_p$.

(4) Calculates the drag force applied on the particle whose top greater than the bed height $H_{bed}$ according to Eq. (8), and performs the aerodynamic entrainment simulations. The criterion for a take-off particle is that the particle mass center is one mean diameter above the bed height.

(5) During the simulation, the bed height is updated according to the accumulated entrained particles as following:

$$H_{bed} = \frac{\sum_{i=1}^{N-N_e} V_{pi}}{(1-p)L_x L_y} \quad (11)$$

where $N$ and $N_e$ are the total and entrained number of bed particles, respectively, and $V_{pi}$ is the volume of particle $i$.

The aerodynamic entrainment processes with various cohesions for both sand and snow are performed, and the simulated cases are shown in Table 2. Each case repeats 5 times considering the randomness of the particle arrangement, and a new sediment bed will be established for each run. Entrained particles will be removed to avoid saltation splashes, and the threshold friction velocity refers to the minimum friction velocity that lifts the first entrained particle.

## 3. Results and discussion

### 3.1 Model validation

The threshold friction velocity determines the onset of the aerodynamic entrainment and subsequently saltation movement. Here, the evolutions of the threshold friction velocity as a function the equivalent diameter under various cohesions are analyzed firstly, which is also compared with existing measurements and predictions. As shown in figure 2a, the obtained threshold friction velocity increases with particle size, consistent with various measurements and predictions. For the same particle size and cohesion, the predicted fluid entrainment threshold may vary within a small range for different sediment beds because of different particle arrangements, which is shown as the error bar (the maximal variation). The cohesion can enhance the critical friction velocity significantly, and the threshold for snow agrees well with the measurements of Clifton *et al.* (2006) when a modest cohesion force is applied. Note that the cohesion force of $F_\Phi$ =50G is on the order of approximately $10^{-5} \sim 10^{-6}$ N, consistent with that measured for snow. Besides, this value generally corresponds to the cohesions in a moist snow sample, because a sintered snow requires much larger cohesions (Comola *et al.*, 2019).

Furthermore, the evolutions of the fluid entrainment threshold versus the inter-particle cohesion for both sand and snow are shown in Figure 3, which are also compared with the prediction model of Shao and Lu (2000). The prediction formula can be read as:

$$u_{*t} = A_N \sqrt{\frac{\rho_p - \rho_f}{\rho_f} g d_p + \frac{\gamma}{\rho_f d_p}}$$

(12)

where $A_N$ is a constant and $\gamma = F_\Phi / d_p$ scales the strength of the inter-particle force. For loose sand and dust, $\gamma$ is on the order of magnitude of approximately $10^{-4}$ Nm$^{-1}$ (Kok and Rennó, 2006; Shao and Lu, 2000).

From Figure 3, the simulated fluid entrainment threshold increases with the inter-particle cohesion force, which agrees well with the prediction. The entrainment over a random sediment bed exhibits strong randomness (see error bars in Figure 2 and 3), which also increases with inter-particle cohesion. The fluid entrainment threshold under larger cohesion may be several times in difference for different sediment beds, mainly because the stress state of exposed particles generally varies drastically under different supporting conditions.

How much time of the first lifting particle takes from a steady state sediment bed (the 'response time') is an interesting but rarely accurately measured values. This also determines whether the duration time of an instantaneous strong fluid shear (i.e., in a turbulent flow) is long enough to entrain surface particles. Figure 4 shows the evolutions of the 'response time' $t_{crit}$ as a function of friction velocity under various cohesions. For snow particles with diameter of 200 $\mu m$, the spending time decreases rapidly (from ~$10^{-2}$ s to ~$10^{-3}$ s) and trends to converge to a threshold of approximately $2.0\times10^{-3}$ s with the increasing friction velocity. The 'response time' also increases with cohesion, and $t_{crit}$ with inter-particle cohesion may several times larger than that without cohesion under the same wind condition.

In a natural turbulent flow, the duration of very-large-scale eddies is comparable with the timescale of the particle initial rolling motion (Cameron *et al.*, 2020). Thus, the 'response time' is much larger than the duration of the largest turbulent structures. The simulations of Dupont *et al.* (2013) also shown that the particle relaxation timescale (comparable with the 'response time') is much larger than the lifetime of the resolved eddies at the near surface.

In order to show the fluid entrainment process of the surface exposed particles intuitively, the entire motion and collision process of several typical entrained particle is also plotted in Figure 5, in which the color indicates the particle resultant velocity in the unit of ms$^{-1}$. Generally speaking, a standing particle starts to move when the fluid drag is large enough to overcome the friction and cohesion, which also accumulates energy from the air flow during its sliding /rolling along the surface. The moving particle may also obtain vertical velocity component by colliding with other raised particles (horizontal momentum transfers to vertical momentum) and result in the aerodynamic lifting in the end. With increasing inter-particle cohesion, surface moving particle requires much larger fluid drag force to initiate the motion, and may experience a longer horizontal distance and more collisions to acquire enough vertical

S. M. Jia and Z. S. Wang

momentum, which is also the main reason for the increasing 'response time' and threshold friction velocity.

*3.2 Entrainment rate*

Always, the entrainment rate is acknowledged to be proportional to the surface shear stress (i.e., $N = \zeta(\tau - \tau_t)$, where $\tau = \rho_f u_*^2$ and $\tau_t = \rho_f u_{*t}^2$ are, respectively, the fluid shear stress at surface and the fluid shear stress threshold, and $\zeta$ is a coefficient with dimensions of $m^2 N^{-1}$) considering the surface momentum balance (Anderson and Haff, 1991). On the one hand, plenty of researches have adopted a friction velocity independent coefficient of $\zeta$ (Doorschot and Lehning, 2002; Zwaaftink *et al.*, 2014), which results in a quadratic dependence of entrainment rate with friction velocity. On the other hand, Shao (2008) have proposed an approximate linear scaling law versus the friction velocity by assuming that the take-off velocity is proportional to the friction velocity, which can be read as:

$$N_e = \xi u_* \left(1 - \frac{u_{*t}^2}{u_*^2}\right) d_p^{-3} \tag{13}$$

where $\xi$ is a dimensionless coefficient that has an order of magnitude around $10^{-3}$.

From comparisons between simulated and fitting curves of Anderson's and Shao's, as shown in Figure 6, it can be seen that Shao's model can describes the tendency of DEM results exactly, and the reason will be discussed in Sec. 3.3. Although inter-particle cohesion increases the fluid entrainment threshold significantly, the entrainment rate seems to be not affected by the cohesion. The reason could be that once particles start to move over the bed surface, the transferred momentum during inter-particle collision is generally large enough to destroy the inter-particle bonds. In this way, the entrainment rate can be described by a uniform function, with only varying fluid entrainment threshold. The fitted curves are also shown in Figure 6, and the parameter $\xi$ is almost one order of magnitude larger than the suggested range, probably because the saltating particle cloud generally acts as a covering of bed grains. However, this simulation neglects this effect by removing the saltating particles.

It is notable that the measurements of Li *et al.* (2020) showed an almost linear scaling of the entrainment rate with the fluid shear stress. This difference could be mainly contributed by the different test settings. For example, DEM simulations remove entrained particles to avoid the saltation splashes, while this strategy is difficult to be implemented in the experiment.

*3.3 Distributions of take-off velocity and angle*

The take-off velocity and angle of entrained particle determines the particle trajectory and subsequently development process of saltation directly. Here, we aim at supplying a valid entrainment scheme from our direct simulations. Except for the entrainment rate given above, the probability density function (PDF) of take-off velocities and angles for snow are shown in Figure 7 and 8, respectively. It can be seen that cohesions do not affect the distributions significantly, thus, we fit the results with and without cohesion using a unified functional form. The unified distribution form is mainly because the bonds of entrained particles have been destroyed during the initial rolling motion. Here, the expressions of lognormal and normal function are given, which will be used in following analyses:

$$prob(\chi) = \frac{1}{\sqrt{2\pi}\sigma_\chi \chi} \exp\left(-\frac{\ln^2(\chi/\mu_\chi)}{2\sigma_\chi^2}\right) \tag{14}$$

$$prob(\chi) = \frac{1}{\sqrt{2\pi}\sigma_\chi} \exp\left(-\frac{(\chi-\mu_\chi)^2}{2\sigma_\chi^2}\right) \tag{15}$$

where $\mu_\chi$ and $\sigma_\chi$ are, respectively, the mean and standard deviation of the distribution.

As shown in Figure 7(a), the normalized take-off velocity (CV) of fluid entrained particle can be described by a lognormal function, with mean and standard deviation expressions are summarized in Table 3. Here, only the results of snow are plotted in the figure, because the distribution pattern is almost the same for both sand and snow. At the same time, the evolution of the mean values with and without cohesion are shown in Figure 7(b). It can be seen that the mean values increase with the friction velocity linearly, which confirm the assumption adopted by (Shao, 2008). However, the take-off velocity of entrained particles is much different from previous assumptions. It's mean value is between the commonly used value of $\sqrt{2gd_p}$ and the value of $3.3u_*$ that does not distinguish entrainments and splash ejections (Zwaaftink *et al.*, 2014).

Furthermore, the distributions of take-off angles are shown in Figure 8, in which $\alpha$ ($\equiv \arctan(w_p/u_p)$) is the take-off angle and $\beta$ ($\equiv \arctan(v_p/u_p)$) is the lateral angle, respectively. It can be found that $\alpha$ and $\beta$ follow the lognormal and normal functions, respectively. Their means and variances are also shown in Table 3. An important feature is that the take-off angle $\alpha$ is much smaller than the commonly used value of $90°$ (leave the bed vertically), and its distribution and mean value is largely consistent with that adopted by (Clifton and Lehning, 2008). Besides, the inter-particle cohesion and friction velocity have no effect on the distributions.

In a word, the cohesion plays an important role in increasing the fluid entrainment threshold, but generally not affects the entrainment rate, take-off velocity and angle distributions. The entrained particles leave the sediment bed with a much smaller take-off angle than $90°$, and generally have modest take-off velocities that falls in between the widely used value of $\sqrt{2gd_p}$ and $3.3\,u_*$ (include splash entrainment). For convenience, the aerodynamic entrainment scheme for sand is also obtained following above steps, which is summarized in table 3.

## 4. Conclusions

Aerodynamic entrainment on a random arranged sediment bed is performed to investigate the entire lifting process of bed particles, in which various cohesions among particles is considered. Air-borne particles generally experience a long distance of rolling/sliding along the bed surface, during which the horizontal momentum transfers to vertical momentum to generate entrainments. The energy accumulated time increases rapidly with inter-particle cohesion, which implies that entrainment on a bonded sediment bed needs the strong fluid shear stress that lasting for a much longer time than that without cohesions.

The take-off velocities and angles are much different from previous assumed values or observations without distinguishing entrainment and splash. The entrainment rate is almost linearly increased as a function the shear velocity, independent of the cohesion, although the cohesion increases the fluid entrainment threshold significantly. The PDFs of take-off velocities and angles are not affected by inter-particle cohesion.

The obtained entrainment scheme largely improves the accurate description of the development process of blown sand and snow, and provides a useful way to explore the role of aerodynamic entrainment within an aeolian transport process, especially within an unsaturated transport.

## Acknowledgements

This work is supported by the National Natural Science Foundation of China (11902341).

## Data statement

The data sets used and/or analyzed during the current study are available from the corresponding author on reasonable request.

## References

Agudo JR, Illigmann C, Luzi G, Laukart A, Delgado A, Wierschem A. 2017. Shear-induced incipient motion of a single sphere on uniform substrates at low particle Reynolds numbers. *Journal of Fluid Mechanics* **825**: 284-314.

Agudo JR, Wierschem A. 2012. Incipient motion of a single particle on regular substrates in laminar shear flow. *Physics of Fluids* **24**(9): 093302.

Anderson RS, Haff PK. 1988. Simulation of Eolian Saltation. *Science* **241**(4867): 820-823.

Anderson RS, Haff PK. 1991. Wind modification and bed response during saltation of sand in air. *Acta. Mech.* **1**: 21-51.

Bagnold RA. 1937. The Transport of Sand by Wind. *The Geographical Journal* **89**: 409-438.

Bagnold RA. 1941. *The Physics of Wind Blown Sand and Desert Dunes*. Methuen: London.

Buffington JM, Montgomery DR. 1997. A Systematic Analysis of Eight Decades of Incipient Motion Studies, With Special Reference to Gravel-Bedded Rivers. *Water Resources Research* **33**: 1993-2029.

Cameron SM, Nikora VI, Witz MJ. 2020. Entrainment of sediment particles by very large-scale motions. *Journal of Fluid Mechanics* **888**: A7.

Carneiro MV, Araújo NA, Pähtz T, Herrmann HJ. 2013. Midair collisions enhance saltation. *Phys.rev.lett* **111**(5): 058001.

Charru F, Larrieu E, Dupont JB, Zenit R. 2007. Motion of a particle near a rough wall in a viscous shear flow. *Journal of Fluid Mechanics* **570**: 431-453.

Chepil WS. 1945. Dynamics of wind erosion .2. Initiation of soil movement. *Soil Science* **60**: 397-411.

Clift R, Grace JR, Weber ME. 1978. *Bubbles, drops, and particles*. Academic Press: New York.

Clifton A, Lehning M. 2008. Improvement and validation of a snow saltation model using wind tunnel measurements. *Earth Surface Processes and Landforms* **33**: 2156-2173.

Clifton A, Rüedi JD, Lehning M. 2006. Snow saltation threshold measurements in a drifting-snow wind tunnel. *Journal of Glaciology* **52**(179): 585-596.

Comola F, Gaume J, Kok JF, Lehning M. 2019. Cohesion-Induced Enhancement of Aeolian Saltation. *Geophysical Research Letters* **46**(10): 5566-5574.

Comola F, Kok JF, Gaume J, Paterna E, Lehning M. 2017. Fragmentation of wind-blown snow crystals. *Geophysical Research Letters* **44**(9): 4195-4203.

Cundall PA. 1987, Distinct element models of rock and soil structure. *Proceedings Analytical and Computational Methods Inn Engineering Rock Mechanics*: 129-163.

Dey S, Papanicolaou A. 2008. Sediment threshold under stream flow: A state-of-the-art review. *Ksce Journal of Civil Engineering* **12**(1): 45-60.

Doorschot JJJ, Lehning M. 2002. Equilibrium Saltation: Mass Fluxes, Aerodynamic Entrainment, and Dependence on Grain Properties. *Boundary-Layer Meteorology* **104**(1): 111-130.

Dupont S, Bergametti G, Marticorena B, Simoëns S. 2013. Modeling saltation intermittency. *Journal of Geophysical Research Atmospheres* **118**(13): 7109-7128.

Dwivedi A, Melville BW, Shamseldin AY, Guha TK. 2011. Flow structures and hydrodynamic force during sediment entrainment. *Water Resources Research* **47**(1): 108-125.

Fletcher B. 1976. The incipient motion of granular materials. *Journal of Physics D* **9**(17): 2471---2478.

Gaume J, van Herwijnen A, Chambon G, Wever N, Schweizer J. 2016. Snow fracture in relation to slab avalanche release: critical state for the onset of crack

...

propagation. *The Cryosphere* **11**: 217-228.

Hofland B, Battjes JA, Booij R. 2005. Measurement of Fluctuating Pressures on Coarse Bed Material. *Journal of Hydraulic Engineering* **131**(9): 770-781.

Houssais M, Ortiz CP, Durian DJ, Jerolmack DJ. 2015. Onset of sediment transport is a continuous transition driven by fluid shear and granular creep. *Nature Communications* **6**: 6527.

Huang N, Shi G. 2017. The significance of vertical moisture diffusion on drifting Snow sublimation near snow surface. *Cryosphere* **11**(6): 3011-3021.

Huang N, Wang Z. 2016. The formation of snow streamers in the turbulent atmosphere boundary layer. *Aeolian Research* **23**: 1-10.

Iversen JD, Pollack JB, Greeley R, White BR. 1976. Saltation threshold on Mars: The effect of interparticle force, surface roughness, and low atmospheric density. *Icarus* **29**(3): 381-393.

Jdoorschot JJ, Lehning M, Vrouwe A. 2004. Field measurements of snow-drift threshold and mass fluxes, and related model simulations. *Boundary-Layer Meteorology* **113**(3): 347-368.

Kok JF, Parteli EJR, Michaels TI, Karam DB. 2012. The physics of wind-blown sand and dust. *Reports on progress in physics. Physical Society (Great Britain)* **75**(10): 106901.

Kok JF, Rennó NO. 2006. Enhancement of the emission of mineral dust aerosols by electric forces. *Geophysical Research Letters* **33**: L19S10.

Lee H, Balachandar S. 2012. Critical shear stress for incipient motion of a particle on a rough bed. *Journal of Geophysical Research Earth Surface* **117**(F1): F01026.

Li G, Zhang J, Herrmann H, Shao Y, Huang N. 2020. Study of Aerodynamic Grain Entrainment in Aeolian Transport. *Geophysical Research Letters* **47**(11): e2019GL086574.

Nemoto M, Nishimura K. 2004. Numerical simulation of snow saltation and suspension in a turbulent boundary layer. *Journal of Geophysical Research Atmospheres* **109**: D18206.

Pähtz T, Clark AH, Valyrakis M, Durán O. 2020. The Physics of Sediment Transport Initiation, Cessation, and Entrainment Across Aeolian and Fluvial Environments. *Reviews of Geophysics* **58**(1): e2019RG000679.

Pähtz T, Durán O. 2020. Unification of Aeolian and Fluvial Sediment Transport Rate from Granular Physics. *Physical Review Letters* **124**(16): 168001.

Pähtz T, Omeradžiˊc A, Carneiro MV, Araújo NAM, Herrmann HJ. 2015. Discrete Element Method simulations of the saturation of aeolian sand transport. *Geophys. Res. Lett.* (42): 2063-2070.

Pan Y, Banerjee S. 1997. Numerical investigation of the effects of large particles on wall-turbulence. *Physics of Fluids* **9**(12): 3786.

Shao Y. 2008. *Physics and Modelling of Wind Erosion*. Springer: Netherlands.

Shao Y, Lu H. 2000. A simple expression for wind erosion threshold friction velocity. *Journal of Geophysical Research Atmospheres* **105**(D17): 22437.

Shih W, Diplas P. 2019. Threshold of Motion Conditions Under Stokes Flow Regime

and Comparison With Turbulent Flow Data. *Water Resources Research* **55**(12): 10872-10892.

Soldati A, Marchioli C. 2009. Physics and modelling of turbulent particle deposition and entrainment: Review of a systematic study. *International Journal of Multiphase Flow* **35**(9): 827-839.

Vinkovic I, Doppler D, Lelouvetel J, Buffat M. 2011. Direct numerical simulation of particle interaction with ejections in turbulent channel flows. *International Journal of Multiphase Flow - INT J MULTIPHASE FLOW* **37**: 187-197.

Vowinckel B, Jain R, Kempe T, Frohlich J. 2016. Entrainment of single particles in a turbulent open-channel flow: a numerical study. *Journal of Hydraulic Research*: 1-14.

White BR. 1982. Two-phase measurements of saltating turbulent boundary layer flow. *International Journal of Multiphase Flow* **8**(5): 459-473.

Yamamoto Y, Potthoff M, Tanaka T, Kajishima T, Tsuji Y. 2001. Large-eddy simulation of turbulent gas-particle flow in a vertical channel: effect of considering inter-particle collisions. *Journal of Fluid Mechanics* **442**(442): 303-334.

Zingg AW. 1953, Wind tunnel studies of the movement of sedimentary material. *Proceedings Proceedings of the Fifth Hydraulic Conference. Studies in Engineering* **34**: 111–135.

Zwaaftink CDG, Diebold M, Horender S, Overney J, Lieberherr G, Parlange MB, Lehning M. 2014. Modelling Small-Scale Drifting Snow with a Lagrangian

Stochastic Model Based on Large-Eddy Simulations. *Boundary-Layer Meteorology* **153**(1): 117-139.

**Figures:**

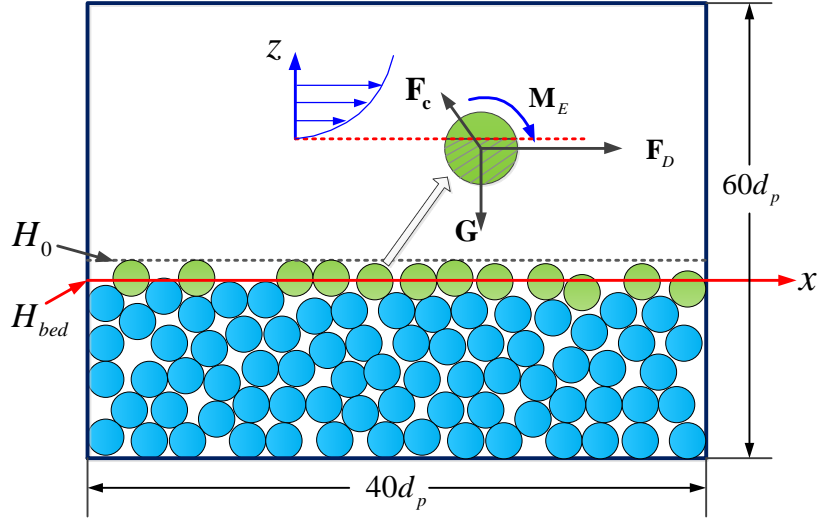

**Figure 1.** Schematic of the determination of the bed height and applied forces on the air-borne particle.

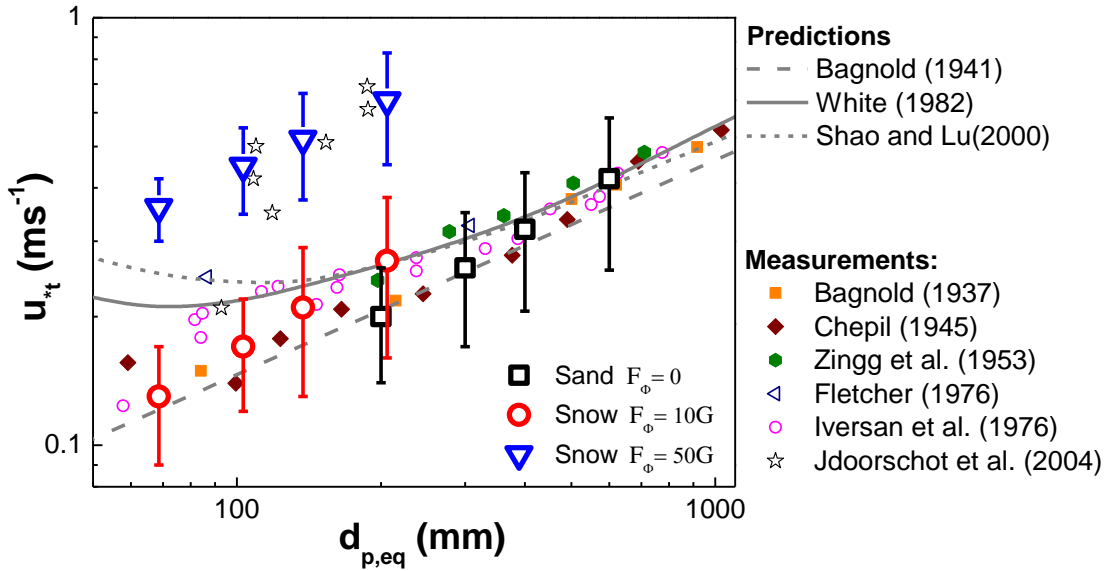

**Figure 2.** Comparisons of simulated entrainment threshold friction velocities ($u_{*t}$) as a function of equivalent particle diameter $d_{p,eq}$ with various predictions and measurements (Bagnold, 1937, 1941; Chepil, 1945; Fletcher, 1976; Iversen *et al.*, 1976; Jdoorschot *et al.*, 2004; Shao and Lu, 2000; White, 1982; Zingg, 1953) (The equivalent particle diameter is defined as $d_{p,eq} = d_p \rho_p / \rho_{sand}$ to consider the correction of the effect of material density).

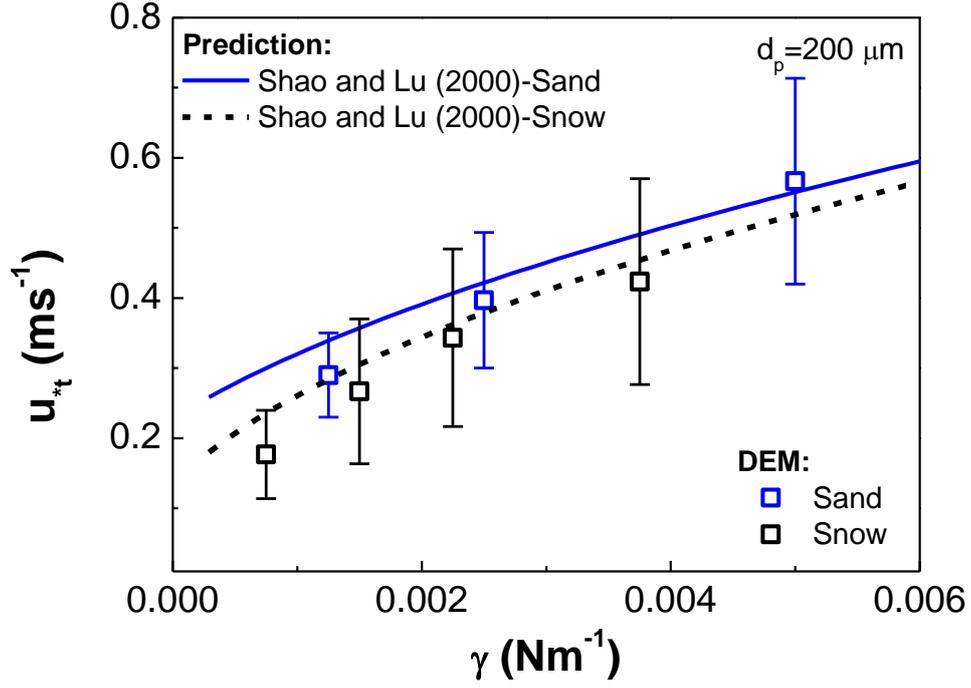

**Figure 3.** Evolutions of the simulated fluid entrainment threshold ($u_{*t}$) versus the inter-particle cohesion (The equivalent cohesion force of the bond is adopted to calculate the parameter $\gamma$ to make DEM results comparable with the prediction).

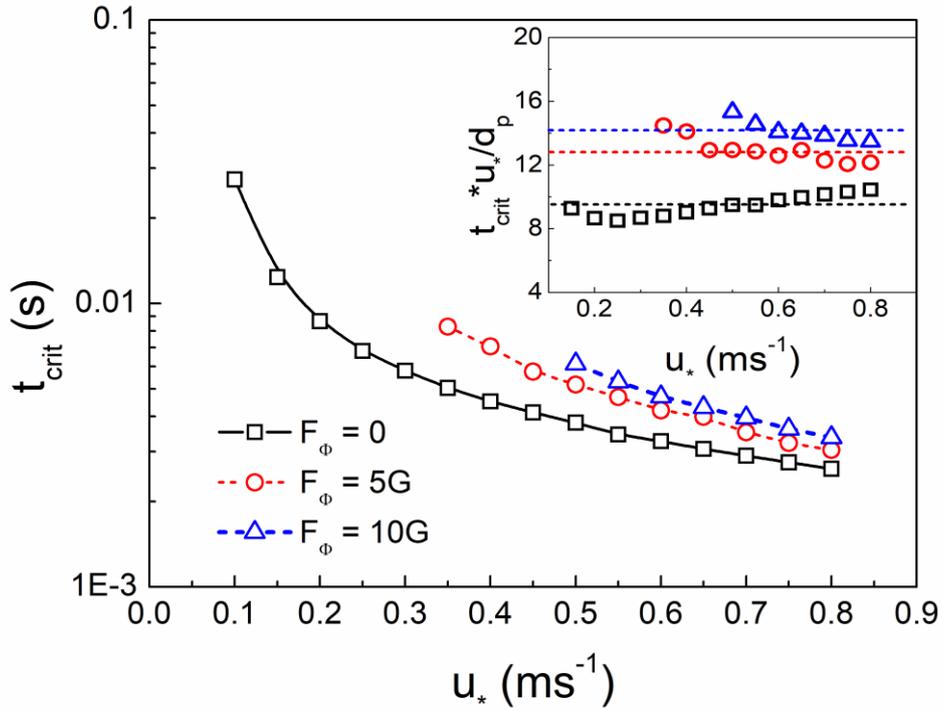

**Figure 4.** Entrainment response time ($t_{cirt}$) as a function of friction velocity under various cohesions for sand.

S. M. Jia and Z. S. Wang

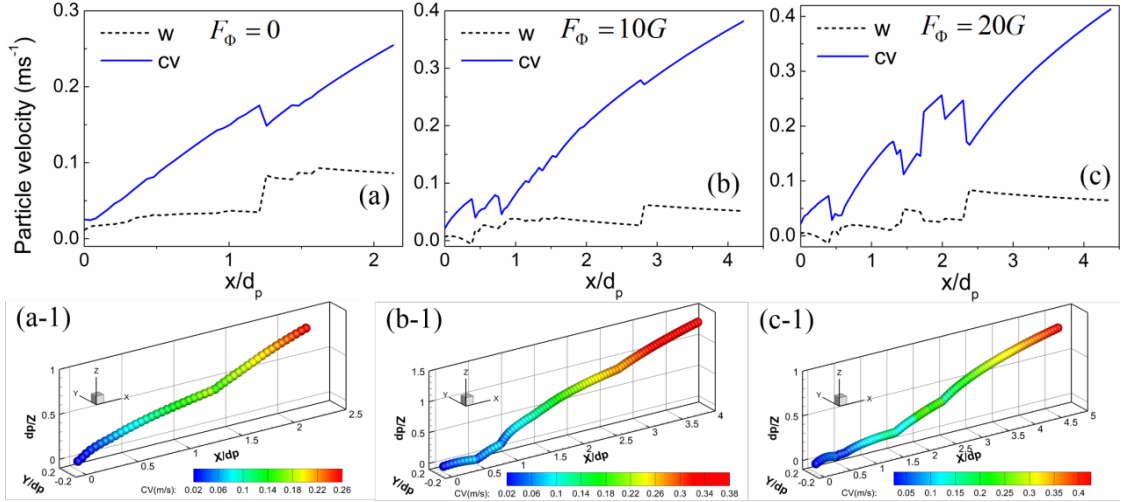

**Figure 5.** Trajectories of typical entrained particles under various cohesions, in which the colors represent the resultant particle velocity.

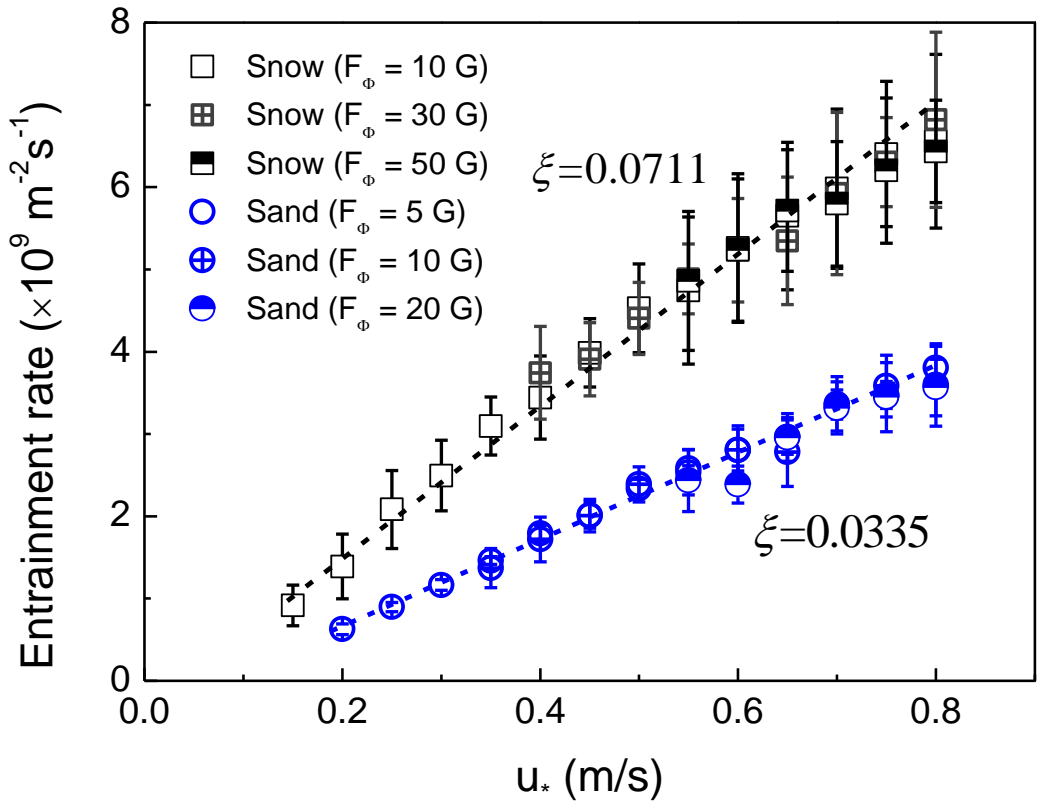

**Figure 6.** Entrainment rate as a function of friction velocities under various cohesions, and dashed lines show the fitted curves using equation (12).

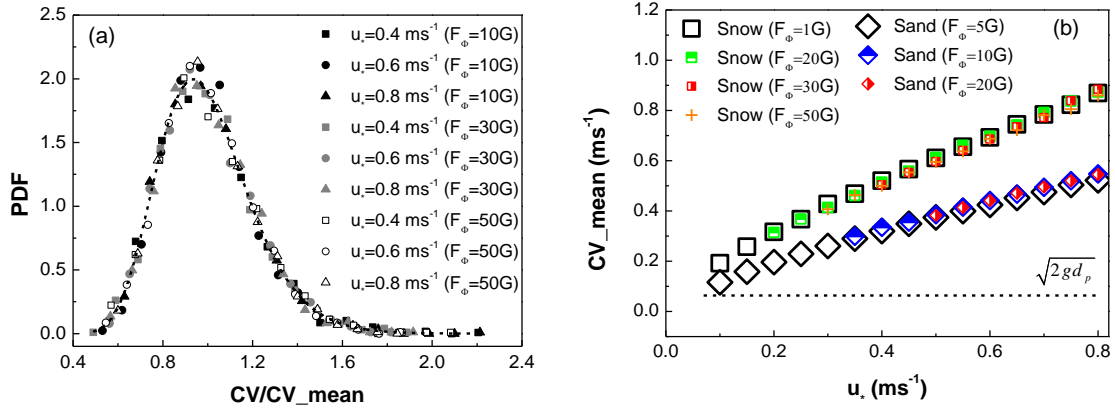

**Figure 7.** (a) Probability density functions (PDF) of the resultant take-off velocities (CV) and (b) variations of the mean take-off velocity versus friction velocity. Here, the black dash lines represent the predictions by $\sqrt{2gd_p}$ (Anderson and Haff, 1991).

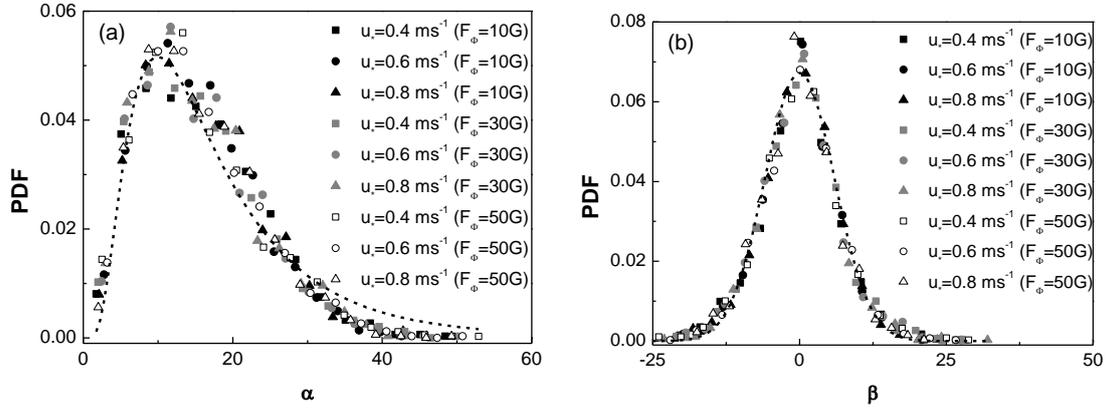

**Figure 8.** PDFs of take-off angles (a) $\alpha$ and (b) lateral $\beta$ for different adhesives $F_\Phi$ and friction velocities $u_*$.

**Tables:**

**Table 1.** Parameters used in DEM simulation.

| Parameter | Description | Value | Unit |
|---|---|---|---|
| $\rho_p$ | Particle density | Snow: 910<br>Sand: 2650 | $kgm^{-3}$ |
| $g$ | Gravitational acceleration | 9.8 | $ms^{-2}$ |
| $\rho$ | Air density | 1.225 | $kgm^{-3}$ |
| $\nu$ | Kinematic viscosity of air | 1.5e-5 | $ms^{-2}$ |
| $k^{*+}$ | Normal-to-shear stiffness ratio | 3.0 | - |
| $\kappa$ | Karman constant | 0.4 | - |
| $\mu^+$ | Frictional coefficient | 0.3 | - |
| $\beta_n^+$ | Normal viscous damping ratio | 0.2 | - |
| $\beta_s^+$ | Tangential viscous damping ratio | 0.2 | - |
| $E^+$ | Young's modulus | $1.0 \times 10^6$ | $Nm^{-2}$ |
| $A_N$ | Entrainment constant | 0.111 | - |

[+] The Young's modulus, viscous damping coefficient and frictional coefficient are chosen following previous DEM simulations (Carneiro *et al.*, 2013; Comola *et al.*, 2019; Pähtz *et al.*, 2015).

**Table 2.** Simulation cases.

| Material | $d_p\ (\mu m)$ | $F_\Phi\ (N)$ |
|---|---|---|
| Sand | 200, 300, 400, 600 | 0~50G |
| Snow | | |

**Table 3.** Aerodynamic entrainment scheme for sand and snow.

| Item | Function | Snow | Sand |
|---|---|---|---|
| $u_{*t}$ | - | $u_{*t} = A_N \sqrt{\dfrac{\rho_p - \rho_f}{\rho_f} g d_p + \dfrac{\gamma}{\rho_f d_p}}$ | |
| $N_e$ | - | $N_e = \xi u_* \left(1 - \dfrac{u_{*t}^2}{u_*^2}\right) d_p^{-3}$ | |
| CV | Lognormal | $\mu_u = 0.13 + 0.95 u_*$ $\sigma_u = 0.21 \mu_u$ | $\mu_u = 0.1 + 0.62 u_*$ $\sigma_u = 0.21 \mu_u$ |
| $\alpha$ | Lognormal | $\mu_\alpha = 14.9°$, $\sigma_\alpha = 0.63°$ | $\mu_\alpha = 15°$, $\sigma_\alpha = 0.74°$ |
| $\beta$ | Normal | $\mu_\beta = 0, \sigma_\beta = 5.9°$ | |